\documentstyle[12pt,graphicx,epsf]{article}
\begin{document}

\def\AEF{A.E. Faraggi}
\def\NPB#1#2#3{{\it Nucl.\ Phys.}\/ {\bf B#1} (#2) #3}
\def\PLB#1#2#3{{\it Phys.\ Lett.}\/ {\bf B#1} (#2) #3}
\def\PRD#1#2#3{{\it Phys.\ Rev.}\/ {\bf D#1} (#2) #3}
\def\PRL#1#2#3{{\it Phys.\ Rev.\ Lett.}\/ {\bf #1} (#2) #3}
\def\PRT#1#2#3{{\it Phys.\ Rep.}\/ {\bf#1} (#2) #3}
\def\MODA#1#2#3{{\it Mod.\ Phys.\ Lett.}\/ {\bf A#1} (#2) #3}
\def\IJMP#1#2#3{{\it Int.\ J.\ Mod.\ Phys.}\/ {\bf A#1} (#2) #3}
\def\nuvc#1#2#3{{\it Nuovo Cimento}\/ {\bf #1A} (#2) #3}
\def\RPP#1#2#3{{\it Rept.\ Prog.\ Phys.}\/ {\bf #1} (#2) #3}
\def\APJ#1#2#3{{\it Astrophys.\ J.}\/ {\bf #1} (#2) #3}
\def\APP#1#2#3{{\it Astropart.\ Phys.}\/ {\bf #1} (#2) #3}
\def\EJP#1#2#3{{\it Eur.\ Phys.\ Jour.}\/ {\bf C#1} (#2) #3}
\def\etal{{\it et al\/}}

\newcommand{\cc}[2]{c{#1\atopwithdelims[]#2}}
\newcommand{\bev}{\begin{verbatim}}
\newcommand{\beq}{\begin{equation}}
\newcommand{\beqa}{\begin{eqnarray}}
\newcommand{\beqn}{\begin{eqnarray}}
\newcommand{\eeqn}{\end{eqnarray}}
\newcommand{\eeqa}{\end{eqnarray}}
\newcommand{\eeq}{\end{equation}}
\newcommand{\Eev}{\end{verbatim}}
\newcommand{\bec}{\begin{center}}
\newcommand{\eec}{\end{center}}
\def\ie{{\it i.e.}}
\def\eg{{\it e.g.}}
\def\half{{\textstyle{1\over 2}}}
\def\nicefrac#1#2{\hbox{${#1\over #2}$}}
\def\third{{\textstyle {1\over3}}}
\def\quarter{{\textstyle {1\over4}}}
\def\m{{\tt -}}
\def\mass{M_{l^+ l^-}}
\def\p{{\tt +}}

\def\slash#1{#1\hskip-6pt/\hskip6pt}
\def\slk{\slash{k}}
\def\GeV{\,{\rm GeV}}
\def\TeV{\,{\rm TeV}}
\def\y{\,{\rm y}}

\def\l{\langle}
\def\r{\rangle}

\begin{titlepage}
\samepage{
\setcounter{page}{1}
\rightline{LTH--742} 
\rightline{arXiv:????.????}
\vspace{1.5cm}
\begin{center}
 {\Large \bf A Novel String Derived $Z^\prime$ With Stable Proton,
Light--Neutrinos and R--parity violation}
\vspace{.25 cm}

 {\large Claudio Corian\`{o}$^{\dagger}$\footnote{ E-mail address:
Claudio.Coriano@le.infn.it},
Alon E. Faraggi$^{\diamondsuit}$\footnote{
E-mail address: faraggi@amtp.liv.ac.uk}
and 
Marco Guzzi$^{\dagger}$\footnote{ E-mail address:
Marco.Guzzi@le.infn.it}
\\
\vspace{.25cm}
{\it $^\dagger$Dipartimento di Fisica, Universita' di Lecce,\\
  I.N.F.N. Sezione di Lecce Via Arnesano, 73100 Lecce, Italy\\}
\vspace{.5cm}
{\it $^\diamondsuit$Department of Mathematical Sciences\\
University of Liverpool, Liverpool, L69 7ZL, United Kingdom}}
\end{center}
\begin{abstract}
The Standard Model indicates the realization of grand unified structures
in nature, and can only be viewed as an effective theory below a higher
energy cutoff. While the renormalizable Standard Model forbids proton
decay mediating operators due to accidental global symmetries,
many extensions of the Standard Model introduce such dimension
four, five and six operators. Furthermore, quantum
gravity effects are expected to induce proton instability, indicating that
the higher energy cutoff scale must be above $10^{16}\GeV$.
Quasi--realistic heterotic string models provide the arena to explore
how perturbative quantum gravity affects the particle physics phenomenology. 
An appealing explanation for the proton longevity is provided
by the existence of an Abelian gauge symmetry that suppresses the
proton decay mediating operators. Additionally, such
a low--scale $U(1)$ symmetry should: allow the suppression
of the left--handed neutrino masses by a seesaw mechanism; allow
fermion Yukawa couplings to the electroweak Higgs doublets; be anomaly
free; be family universal. These requirements render the existence of
such $U(1)$ symmetries in quasi--realistic heterotic string models 
highly non--trivial. We demonstrate the existence of a $U(1)$ symmetry
that satisfies all of the above requirements in a class of left--right
symmetric heterotic string models in the free fermionic formulation. 
The existence of the extra $Z^\prime$ in the energy range accessible to 
future experiments is motivated by the requirement of adequate
suppression of proton decay mediation. We further show that while 
the extra $U(1)$ forbids dimension four 
baryon number violating operators it allows dimension four
lepton number violating  operators and $R$--parity violation.

\end{abstract}
\smallskip}
\end{titlepage}

\section{Introduction}

The Standard Model of particle physics successfully accounts for 
all observations in the energy range accessible to
contemporary experiments. Despite this enormous success the Standard
Model can only be viewed as an effective low energy field theory 
below a higher energy cutoff. In the least, the existence of a Landau pole in
the hypercharge sector, albeit at an enormously high scale, unequivocally
demonstrates the formal inconsistency of the Standard Model.
In this regard, the renormalizability of the Standard Model
is an approximate feature and effects of nonrenormalizable
operators, suppressed by powers of the high scale cutoff, 
must be considered.

The high precision analysis of the Standard Model parameters,
achieved at LEP and other particle physics experiments,
indicates that the Standard Model remains an approximate
renormalizable quantum field theory up to a very large 
energy scale. Possibly the grand unification
scale, or the Planck scale. The logarithmic evolution of the Standard 
Model parameters is in agreement with the available data,
and is compatible with the notion
of unification at a high energy scale in the gauge and heavy
matter sectors of the Standard Model. Preservation
of the logarithmic evolution in the scalar sector
necessitates the introduction of a new 
symmetry between bosons and fermions, dubbed supersymmetry.

Perhaps the most important observation indicative that the 
Standard Model cutoff scale is a very high scale is the 
longevity of the proton. Renormalizability insures that 
baryon and lepton violating operators
are absent in the perturbative Standard Model.
Hence, in the renormalizable Standard Model
baryon and lepton numbers are accidental global symmetries. 
However, at the cutoff scale dimension six operators are 
induced and the proton is in general expected to decay.
The observed proton lifetime implies that the cutoff scale
is of order $10^{16}$GeV. 
The problem is exacerbated in supersymmetric extensions
of the Standard Model that allow dimension four and five
baryon and lepton violating operators \cite{wei}.
Indeed, one would expect proton decay mediating operators
to arise in most extensions of the Standard Model.
In the Minimal Supersymmetric Standard Model one imposes
a global symmetry, $R$--parity, which forbids the
dimension four baryon and lepton number violating operators.
The difficulty with dimension five operators can only 
be circumvented if one further assumes that the relevant 
Yukawa couplings are suppressed. However, as global symmetries
are not expected to survive quantum gravity effects \cite{qge},
the proton lifetime problem becomes
especially acute in the context of theories that 
unify the Standard Model with gravity. 
This question has
been examined extensively in the context of quasi--realistic 
heterotic string models. In this context, the most
appealing suggestion is that the suppression of the
proton decay mediating operators is a result of
a gauged $U(1)$ symmetry, under which the undesired
nonrenormalizable dimension four and five operators are
not invariant. If the $U(1)$ symmetry remains unbroken down
to sufficiently low scales the problematic operators will be 
suppressed by at least the VEV that breaks the additional 
$U(1)$ symmetry over the cutoff scale. 

The free fermionic heterotic string models are
among the most realistic string models constructed to date
\cite{fsu5,fny,alr,slm,lrs,su421}.
The issue of proton stability was sporadically explored in these
models \cite{lepzprime,pstudies, pati, ps, thor}, 
as well as explorations of possible $U(1)$ symmetries
that can ensure proton longevity \cite{lepzprime, pati, ps, thor}.
However, non of the current proposals is satisfactory.
The $U(1)$ symmetry of ref. \cite{lepzprime} is the $U(1)$
combination of $B-L$ and $T_{3_R}$ which is embedded in $SO(10)$
and is orthogonal to the electroweak hypercharge. However, this
$U(1)$ symmetry in general needs to be broken to allow for
the suppression of the left--handed neutrino masses by a seesaw mechanism. 
Similarly, the $U(1)$ symmetries studied in ref. \cite{pati,ps,thor},
that arise in the string models from combinations of the $U(1)$
symmetries that are external to $SO(10)$ are flavour dependent $U(1)$
symmetries that in general must be broken near the string scale to 
allow for generation of fermion masses. In ref. \cite{ps}
it was concluded that non of the symmetries suggested in
ref. \cite{pati} can remain unbroken down to low energies and provide
for the suppression of the proton decay mediating operators.
Furthermore, a family non--universal $U(1)$ symmetry is 
restricted by constraints on flavour changing neutral currents,
and cannot exist in energy range accessible to forthcoming experiments.

The proton longevity, together with the Standard Model multiplet structure,
therefore provide the most important clues for the origin of the 
Standard Model particle spectrum. These favour the
embedding of Standard Model in a Grand Unified Theory, possibly
broken to the Standard Model at the string level.
The GUT embedding of the Standard Model, and its supersymmetric extension, 
leads to proton decay mediating operators.
The most robust and economical way to suppress the dangerous
operators is by the existence of an additional Abelian gauge
symmetry which is broken above the electroweak scale and
does not interfere with the other phenomenological constraints.
Such a $U(1)$ symmetry should fulfill the following requirements:
\begin{itemize}
\item Forbid dimension four, five and six proton decay
mediating operators.
\item Allow suppression of left--handed neutrino masses
by a seesaw mechanism.
\item Allow the fermion Yukawa couplings to electroweak Higgs
doublets.
\item Be family universal.
\item Be anomaly free.
\end{itemize}

This list of requirements render the existence of such a 
$U(1)$ symmetry in string models highly nontrivial. For example,
in models with an underlying $SO(10)$ GUT embedding the $U(1)_{B-L}$
symmetry is gauged. It forbids the dimension four baryon and 
lepton number violating operators, but not the dimension five
operator. Furthermore, suppression of left--handed neutrino masses
by a seesaw mechanism in general necessitates that the symmetry 
is broken near the GUT scale. Hence, it cannot remain 
unbroken down to low energies, and in general fast
proton decay from dimension four operators is expected to ensue. 
Similarly, the $U(1)_A$ symmetry external to $SO(10)$ in
$E_6\rightarrow SO(10)\times U(1)_A$ is
anomalous in many of the quasi--realistic 
string models constructed to date \cite{cleaverau1}
and is broken by a generalised
Green--Schwarz mechanism. The additional $U(1)$s investigated 
in refs. \cite{pati,ps,thor} are either flavour non--universal
or constrain the fermion Yukawa mass terms and must therefore
be similarly broken near the Planck scale. Thus, of all the 
extra $U(1)$'s investigated to date non seems to remain viable
down to low energies, and to provide the coveted proton protection
symmetry.

In this paper we therefore explore further the possibility that 
quasi--realistic string models give rise to Abelian gauged symmetries
that can play the role of the proton lifetime guard. 
We demonstrate the existence of a $U(1)$ symmetry satisfying all of
the above requirements in the class of 
left--right symmetric string--derived models of ref. \cite{lrs}.
The key to obtaining the $U(1)$ symmetry satisfying the 
above requirements is the $SO(10)$ symmetry breaking pattern 
particular to the left--right symmetric models \cite{lrs}.
The key distinction is that in these models the
$U(1)_A$, which is external to the unbroken $SO(10)$ subgroup,
is anomaly free, and may remain unbroken down to low energies. 
It is does not restrict the charged fermion mass
terms, and it allows for the suppression of the left--handed neutrino
masses by a seesaw mechanism. 
Its existence at low energies is 
motivated by the longevity of the proton lifetime. 
Furthermore, as we discuss below, 
while it forbids the supersymmetric dimension four and five
baryon number violating operators,
it allows the dimension four lepton number violating operator.
Hence, while proton decay from dimension four operators does not ensue,
lepton number and $R$--parity violation do arise.
This observation 
has far reaching implications in terms of the phenomenology and
collider signatures of the models.

\section{The structure of the free fermionic models}

In this section we describe the structure of the quasi--realistic 
free fermionic models and the properties of the proton
protecting $U(1)$ symmetry. 
The free fermionic formulation
the 4-dimensional heterotic string,
in the light-cone gauge, is described
by $20$ left--moving  and $44$ right--moving two dimensional real
fermions \cite{fff}.
The models are constructed by specifying
the phases picked up by the world--sheet fermions 
when transported around the torus non-contractible loops.
Each model corresponds to a particular choice of 
fermion phases consistent with
modular invariance
that can be generated by a set of basis vectors $v_i,i=1,\dots,n$,
$v_i=\left\{\alpha_i(f_1),\alpha_i(f_{2}),\alpha_i(f_{3}))\dots\right\}.$
The basis vectors span a space $\Xi$ which consists of $2^N$ sectors that give
rise to the string spectrum.
The spectrum is truncated by a
Generalised GSO (GGSO) projections \cite{fff}.

The $U(1)$ charges, $Q(f)$, with respect to the unbroken Cartan
generators of the four 
dimensional gauge group, which are in one 
to one correspondence with the $U(1)$
currents ${f^*}f$ for each complex fermion $f$, are given by:
\beqn
{Q(f) = {1\over 2}\alpha(f) + F(f)},
\label{u1charges}
\eeqn
where $\alpha(f)$ is the boundary condition of the world--sheet fermion $f$
in the sector $\alpha$.
$F(f)$ is the fermion number operator counting each mode of 
$f$ once (and if $f$ is complex, $f^*$ minus once). 
For periodic fermions,
$\alpha(f)=1$, the vacuum is a spinor in order to represent the Clifford
algebra of the corresponding zero modes. 
For each periodic complex fermion $f$
there are two degenerate vacua ${\vert +\rangle},{\vert -\rangle}$ , 
annihilated by the zero modes $f_0$ and
${{f_0}^*}$ and with fermion numbers  $F(f)=0,-1$, respectively. 

The two dimensional world--sheet fermions are divided 
in the following way:
the eight left--moving real fermions $\psi^{1,2}$ and
$\chi^{1,\cdots,6}$ correspond to the eight Ramond--Neveu--Schwarz
fermions of the ten dimensional heterotic string in the light--cone 
gauge; the twenty--four real--fermions
$\{y^i, \omega^i\vert \bar{y}^i,\bar{\omega}^i\}$, 
$i=1,\dots,6$ correspond to the fermionized internal 
coordinates of a compactified manifold in a bosonic formulation;
the complex right--moving fermions
${\bar\phi}^{1,\cdots,8}$ generate the rank eight hidden gauge group; 
${\bar\psi}^{1,\cdots,5}$ generate the $SO(10)$ gauge group;
${\bar\eta}^{1,2,3}$ generate the three remaining $U(1)$
generators in the Cartan sub-algebra of
the observable rank eight gauge group.
A combination of these $U(1)$ currents will play the role of
the proton lifetime guard. 

The free fermionic models are defined in terms of the
basis vectors and one--loop GGSO projection coefficients.
The quasi--realistic free fermionic heterotic--string model
are typically constructed in two stages.
The first stage consists of the NAHE--set, 
$\{{\bf 1}, S,b_1,b_2,b_3\}$ \cite{costas, nahe}.
The gauge group at this stage is $SO(10)\times SO(6)^3\times E_8$,
and the vacuum contains forty--eight multiplets in the 16 chiral
representation of $SO(10)$. The second stage consists of adding three
or four basis vectors to the NAHE--set, typically denoted by
$\{\alpha,\beta,\gamma\}$. The additional basis vectors
reduce the number of generations to three, with one arising
from each of the basis vectors $b_1$, $b_2$ and $b_3$. Additional
non--chiral generations may arise from the basis vectors
that extend the NAHE--set. 
This distribution of the chiral generations is particular
to the class of quasi--realistic free fermionic models that
has been explored to date, and other possibilities may exist \cite{fkrII}. 
Additionally, the basis vectors that extend the NAHE--set
break the four dimensional gauge group. The $SO(10)$ symmetry 
is broken to one of the subgroups: $SU(5)\times U(1)$ \cite{fsu5};
$SO(6)\times SO(4)$ \cite{alr}; $SU(3)\times SU(2)\times U(1)^2$
\cite{slm}; $SU(3)\times SU(2)^2\times U(1)$ \cite{lrs}; or 
$SU(4)\times SU(2)\times U(1)$ \cite{su421}. 
The three generations
from the sectors $b_1$, $b_2$ and $b_3$ are decomposed under the
final $SO(10)$ subgroup. 
The flavour $SO(6)^3$ groups are broken to products of $U(1)^n$ with
$3\le n\le 9$. The $U(1)^{1,2,3}$ factors arise from the three 
right--moving complex fermions ${\bar\eta}^{1,2,3}$. Additional $U(1)$ 
currents may arise from complexifications of right--moving fermions from the 
set $\{{\bar y},{\bar\omega}\}^{1,\cdots,6}$. 

The $U(1)$ symmetry that will serve as the proton lifetime guard
is a combination of the three $U(1)$ symmetries generated by the 
world--sheet complex fermions ${\bar\eta}^{1,2,3}$. The states from 
each of the sectors $b_1$, $b_2$ and $b_3$ are charged with respect
to one of these $U(1)$ symmetries, {\it i.e.} with respect to 
$U(1)_1$, $U(1)_2$ and $U(1)_3$, respectively. Hence the $U(1)$ combination
\beq
U(1)_\zeta=U_1+U_2+U_3
\label{u1zeta}
\eeq
is family universal. In the string derived models of ref.
\cite{fsu5,fny,alr,slm} $U(1)_{1,2,3}$ are anomalous. 
Therefore, also $U(1)_\zeta$ is anomalous and must be
broken near the string scale. In the string derived left--right
symmetric models of ref \cite{lrs} $U(1)_{1,2,3}$ are anomaly free,
and hence also the combination $U(1)_\zeta$ is anomaly free.
It is this property of these models which allows this 
$U(1)$ combination to remain unbroken. 

Subsequent to constructing the basis vectors and extracting the massless
spectrum the analysis of the free fermionic models proceeds by
calculating the superpotential. 
The cubic and higher-order terms in the superpotential 
are obtained by evaluating the correlators
\beq
A_N\sim \langle V_1^fV_2^fV_3^b\cdots V_N\rangle,
\label{supterms}
\eeq
where $V_i^f$ $(V_i^b)$ are the fermionic (scalar) components
of the vertex operators, using the rules given in~\cite{kln}.
Generically, correlators of the form (\ref{supterms}) are of order
${\cal O} (g^{N-2})$, and hence of progressively higher orders
in the weak-coupling limit.
Typically, 
one of the $U(1)$ factors in the free-fermion models is anomalous,
and generates a Fayet--Ilioupolos term which breaks supersymmetry
at the Planck scale \cite{dsw}.
The anomalous $U(1)$ is broken, and supersymmetry
is restored, by a non--trivial VEV for some scalar
field that is charged under the anomalous $U(1)$.
Since this field is in general also charged with respect
to the other anomaly-free $U(1)$ factors, some non-trivial
set of other fields must also get non--vanishing VEVs $\cal V$,
in order to ensure that the vacuum is supersymmetric.
Some of these fields will appear in the nonrenormalizable terms
(\ref{supterms}), leading to
effective operators of lower dimension. Their coefficients contain
factors of order ${\cal V} / M{\sim 1/10}$.
Typically the solution of the D-- and F--flatness
constraints break most or all of the horizontal $U(1)$ symmetries.

\section{The proton lifeguard}
In this section we discuss the characteristics of $U(1)_\zeta$ in the 
left--right symmetric string derived models \cite{lrs}, versus those
of $U(1)_A$ in the string derived models of refs. \cite{fsu5,fny,alr,slm}.
We note that both $U(1)_\zeta$ as well as $U(1)_A$
are obtained from the same combination of complex right--moving 
world--sheet currents $\bar\eta^{1,2,3}$, {\it i.e.} both are 
given by a combination of $U_1$, $U_2$, and $U_3$.
The distinction between the two cases, as we describe in detail below,
is due to the charges of
the Standard Model states, arising from the sectors $b_1$, $b_2$ and $b_3$,
under this combination.
The key feature of $U(1)_\zeta$ in the models of ref. \cite{lrs} is that it
is anomaly free.
To study the characteristics of the proton protecting
$U(1)$ symmetry it is instructive
to examine in combinatorial notation 
the vacuum structure of the chiral generations 
from the sectors $b_{1,2,3}$.
The vacuum of the sectors
$b_j$ contains twelve periodic fermions. Each periodic fermion
gives rise to a two dimensional degenerate vacuum $\vert{+}\rangle$ and
$\vert{-}\rangle$ with fermion numbers $0$ and $-1$, respectively.
The GSO operator, is a generalised parity operator, which
selects states with definite parity. After applying the
GSO projections, we can write the degenerate vacuum of the sector
$b_1$ in combinatorial form
\begin{eqnarray}
{\left[\left(\matrix{4\cr
                                    0\cr}\right)+
\left(\matrix{4\cr
                                    2\cr}\right)+
\left(\matrix{4\cr
                                    4\cr}\right)\right]
\left\{\left(\matrix{2\cr
                                    0\cr}\right)\right.}
&{\left[\left(\matrix{5\cr
                                    0\cr}\right)+
\left(\matrix{5\cr
                                    2\cr}\right)+
\left(\matrix{5\cr
                                    4\cr}\right)\right]
\left(\matrix{1\cr
                                    0\cr}\right)}\nonumber\\
{+\left(\matrix{2\cr
                                    2\cr}\right)}
&{~\left[\left(\matrix{5\cr
                                    1\cr}\right)+
\left(\matrix{5\cr
                                    3\cr}\right)+
\left(\matrix{5\cr
                                    5\cr}\right)\right]\left.
\left(\matrix{1\cr
                                    1\cr}\right)\right\}}
\label{spinor}
\end{eqnarray}
where
$4=\{y^3y^4,y^5y^6,{\bar y}^3{\bar y}^4,
{\bar y}^5{\bar y}^6\}$, $2=\{\psi^\mu,\chi^{12}\}$,
$5=\{{\bar\psi}^{1,\cdots,5}\}$ and $1=\{{\bar\eta}^1\}$.
The combinatorial factor counts the number of $\vert{-}\rangle$ in the
degenerate vacuum of a given state.
The first term in square brackets counts the degeneracy of the
multiplets, being eight in this case. 
The two terms in the curly brackets correspond to the two
CPT conjugated components of a Weyl spinor. The first
term among those corresponds to the 16 spinorial representation
of $SO(10)$, and fixes the space--time chirality properties of the
representation, whereas the second corresponds to the CPT conjugated
anti--spinorial $\overline{16}$ representation. 
Similar vacuum structure is obtained
for $b_2$ and $b_3$. The periodic boundary conditions of the world--sheet
fermions ${\bar\eta}^j$ entails that the fermions from each sector
$b_j$ are charged with respect to one of the $U(1)_j$ symmetries. 
The charges, however, depend on the $SO(10)$ symmetry breaking 
pattern, induced by the basis vectors that extend the NAHE--set, and
may, or may not, differ in sign between different components of a given
generation. In the models of ref. \cite{fsu5, slm, alr} 
the charges of a given $b_j$ generation under $U(1)_j$ is of the 
same sign, whereas in the models of ref. \cite{lrs} they differ.
In general, the distinction is by the breaking of $SO(10)$ to
either $SU(5)\times U(1)$ or $SO(6)\times SO(4)$. In the former case
they will always have the same sign, whereas in the later they may
differ. This distinction fixes the charges of the Standard 
Model states under the $U(1)$ symmetry which safeguards the proton
from decaying, while not obstructing the remaining constraints
listed above. 

In the free fermionic standard--like models the $SO(10)$ symmetry
is broken to\footnote{
$U(1)_C=3/2 U(1)_{B-L}~;~U(1)_L=2 U(1)_{T_{3_R}}$}
$SU(3)\times SU(2)\times U(1)_C\times U(1)_L$.
The weak hypercharge is given by 
\beq
U(1)_Y={1\over3} U(1)_C+{1\over2}U(1)_L, \label{u1y}
\eeq
and the orthogonal $U(1)_{Z^\prime}$ combination is given 
by 
\beq
U(1)_{Z^\prime}=U(1)_C-U(1)_L. \label{u1zprime}
\eeq

The three twisted 
sectors $b_1$, $b_2$ and $b_3$ produce three generations
in the sixteen representation of $SO(10)$ decomposed 
under the final $SO(10)$ subgroup. In terms of the
$SU(3)_C\times U(1)_C\times SU(2)_L\times U(1)_L$
decomposition they take the values
\beqn
{E}&&\equiv ~[(      1 ,~~{3/2});(1,~~1)];\nonumber\\
{U}&&\equiv ~[({\bar 3}, -{1/2});(1, -1)];\nonumber\\
 Q &&\equiv ~[(      3 ,~~{1/2});(2,~~0)];\nonumber\\
{N}&&\equiv ~[(      1 ,~~{3/2});(1, -1)];\nonumber\\
{D}&&\equiv ~[({\bar 3}, -{1/2});(1,~~1)];\nonumber\\
 L &&\equiv ~[(      1 , -{3/2});(2,~~0)].
\label{su3211decomposition}
\eeqn

In terms of the $SO(6)\times SO(4)$ Pati--Salam decomposition
\cite{patisalam} the Standard Model fermion fields are embedded in the
\beqn
{F_L}&&\equiv ~(      4 ,2,1)=~~~Q~~~+~~~L~~;\nonumber\\
{F_R}&&\equiv ~({\bar 4},1,2)=U+D + E+N~~,
\label{so64decomposition}
\eeqn
representations of $SU(4)\times SU(2)_L\times SU(2)_R$. In terms
of the left--right symmetric decomposition of ref. \cite{lrs}
the embedding is in the following
representations:
\beqn
Q_L &=& (      3 ,2,1,~{~{1\over2}})~~,  \\
Q_R &=& ({\bar 3},1,2, {-{1\over2}}) ~=~  {U+D}~~,\\
L_L &=& (      1 ,2,1, {-{3\over2}}) ~~,\\
L_R &=& (      1 ,1,2,~{~{3\over2}}) ~~=~  {E+N}~~,
\label{LRsymreps}
\eeqn
of $SU(3)\times SU(2)_L\times SU(2)_R\times U(1)_C$.
The Higgs fields in the later case are in a bi--doublet representation
\beq
h       = (1,2,2,0) ~~=~  
{\left(\matrix{
                 h^u_+  &  h^d_0\cr
                 h^u_0  &  h^d_-\cr}\right)}~~. 
\label{HiggsLRsymreps}
\eeq

Using the combinatorial notation introduced in eq. (\ref{spinor})
the decomposition of the 16 representation of $SO(10)$ in the 
Pati--Salam string models is

\beq
\{\left[\left(\matrix{3\cr
                                    0\cr}\right)+
\left(\matrix{3\cr
                                    2\cr}\right)\right]
\left[\left(\matrix{2\cr
                                    0\cr}\right)+
\left(\matrix{2\cr
                                    2\cr}\right)\right]\}+
\{\left[\left(\matrix{3\cr
                                    1\cr}\right)+
\left(\matrix{3\cr
                                    3\cr}\right)\right]
\left[\left(\matrix{2\cr
                                    1\cr}\right)\right]\}
\label{psspinor}
\eeq
The crucial point is that the Pati--Salam breaking pattern
allows the first and second terms in curly brackets to 
come with opposite charges under $U(1)_j$. This results from the 
operation of the GSO projection operator, which differentiates
between the two terms. Thus, in models that descend from $SO(10)$
via the $SU(5)\times U(1)$
breaking pattern the charges of a generation from a sector
$b_j$ $j=1,2,3$, under the corresponding symmetry $U(1)_j$ 
are either $+1/2$, or $-1/2$, for all the states from that sector.
In contrast, in the left--right symmetric string models the
corresponding charges, up to a sign are, 
\beq
Q_j(Q_L;L_L)=+1/2~~~;Q_j(Q_R;L_R)=-1/2,
\label{u1chargesinlrsmodel}
\eeq
{\it i.e.} the charges of the $SU(2)_L$ doublets have the opposite 
sign from those of the $SU(2)_R$ doublets. 
This is in fact the reason that in the left--right symmetric string models
\cite{lrs}
it was found that, in contrast to the case of the FSU5 \cite{fsu5}, 
Pati--Salam \cite{alr} and standard--like \cite{slm},
string models, the $U(1)_j$ symmetries are not part of the anomalous
$U(1)$ symmetry \cite{lrs}.

It is therefore noted that the 
\beq
U(1)_\zeta=U_1+U_2+U_3
\eeq
combination is a family--universal,
anomaly free\footnote{We note that there may exist
string models in the classes of \cite{fsu5, alr, slm} in
which $U(1)_\zeta$ is anomaly free.
This may be the case in the so called self--dual vacua of ref. \cite{fkrII}.
Such quasi--realistic string models with an anomaly free $U(1)_\zeta$ have
not been constructed to date.}, $U(1)$ symmetry, and
allows the quark and lepton fermion mass terms 
\beq
Q_LQ_Rh~~~~~{\rm and}~~~~~L_LL_Rh~.
\eeq
The two combinations of $U(1)_1$, $U(1)_2$ and $U(1)_3$,
that are orthogonal to $U(1)_\zeta$, are family non--universal
and may be broken at, or slightly below, the string scale.

The left--right symmetric heterotic string models of ref. \cite{lrs}
provide explicit quasi--realistic string models, that realize the
charge assignment of eq. (\ref{u1chargesinlrsmodel}).
Furthermore, the dimension four and five baryon number violating operators
that arise from 
\beqn
{Q_LQ_LQ_LL_L} & \rightarrow & QQQL \label{qqql}\\
{Q_RQ_RQ_RL_R} & \rightarrow &  \{UDDN,UUDE\} \label{uddn}
\eeqn
are forbidden, while the lepton number violating operators
that arise from
\beqn
{Q_LQ_RL_LL_R} & \rightarrow & QDLN \label{qdln}\\
{L_LL_LL_RL_R} & \rightarrow & LLEN \label{llen}
\eeqn
are allowed. 

The crucial observation is the opposite charge assignment 
of the left and right--handed fields under $U(1)_\zeta$. This
is available in models that descend from the Pati--Salam symmetry
breaking pattern of the underlying $SO(10)$ GUT symmetry. 
In this case the left-- and right--moving fields carry opposite sign
under the GSO projection operator, induced by the basis 
vector that breaks $SO(10)\rightarrow SO(6)\times SO(4)$. 
An additional symmetry breaking stage of the Pati--Salam models \cite{alr},
or left--right symmetric models \cite{lrs},
can be obtained at the string level or in the effective 
low--energy field theory by the Higgs fields in the representations
$\{Q_H,{\bar Q}_H\}=\{({\bar 4},1,2), (4,1,2)\}$
or 
$\{L_H,{\bar L}_H\}=\{({\bar 1},1,2,{3\over2}), (1,1,2,{-{3\over2}})\}$.
The breaking can be achieved at the string level, while preserving the
desired charge assignment, as long as a basis vector of the
form $2\gamma$ of refs. \cite{slm}, or $b_6$ of ref. \cite{alr},
are not introduced. The boundary condition assignments in these 
basis vectors entails that the $N=4$ vacuum that we start with factorizes
the gauge degrees of freedom into $E_8\times E_8$ or $SO(16)\times SO(16)$.
The consequence of this is that all the states from the twisted
matter sectors $b_j$ carry the same charge under $U(1)_j$. Thus,
this result is circumvented by not including the vectors
$2\gamma$ of \cite{slm}, or $b_6$ of \cite{alr} in the construction.
In effect, such models are descending from a different $N=4$ 
underlying vacuum \cite{lrs,su421}. Being $SO(16)\times E_7\times E_7$
in the models of ref. \cite{lrs}, which 
explicitly realize the desired breaking pattern in a class of 
quasi--realistic string models. 
We assume below that the $SU(2)_R$
symmetry is broken directly at the string level
in which case the remnant $U(1)_{Z^\prime}$ given in eq. (\ref{u1zprime})
has to be broken by the Higgs fields $\{ N_H,{\bar N}_H\}=
(1,1,0,5/2), (1,1,0,-5/2)$ under
$SU(3)\times SU(2)\times U(1)_Y\times U(1)_{Z^\prime}$.

\section{An effective string inspired $Z^\prime$ model}
 
Inspired by the $U(1)$ charge assignment in the left--right symmetric
string derived models \cite{lrs}, we present an effective field theory
model incorporating these features. At this stage our aim is 
to build an effective model that can be used in correspondence
with experimental data, rather than a complete effective field theory
model below the string scale, which is of further interest and will be
discussed in future publications.
The charges of the fields in the 
low energy effective field theory of the string inspired model
are given by 

\beqn
\begin{tabular}{|c|rrrr|}
\hline
\bf{Field}&$U(1)_Y$&$U(1)_{Z'}$&$U(1)_{\zeta}$&$U(1)_{\zeta^\prime}$\\
\hline
${Q}^i$ & $\frac{1}{6}$ &$ \frac{1}{2}$& $-\frac{1}{2}$& $ \frac{3}{5}$ \\
${L}^i$ &$-\frac{1}{2}$ &$-\frac{3}{2}$& $-\frac{1}{2}$& $ \frac{1}{5}$  \\
${U}^i$ &$-\frac{2}{3}$ &$ \frac{1}{2}$& $ \frac{1}{2}$& $-\frac{2}{5}$  \\
${D}^i$ & $\frac{1}{3}$ &$-\frac{3}{2}$& $ \frac{1}{2}$& $-\frac{4}{5}$  \\
${E}^i$ & $      1    $ &$ \frac{1}{2}$& $ \frac{1}{2}$& $-\frac{2}{5}$  \\
${N}^i$ & $      0    $ &$ \frac{5}{2}$& $ \frac{1}{2}$& $     0      $  \\
$\phi^i$ &$      0    $ &$ 0          $& $      0     $& $     0      $  \\
$\phi^0$ &$      0    $ &$ 0          $& $      0     $& $     0      $  \\
$H^U$ & $ \frac{1}{2}$&$     -1     $& $    0         $& $-\frac{1}{5}$   \\
$H^D$ & $-\frac{1}{2}$&$     1      $& $    0         $& $ \frac{1}{5}$  \\
$N_H$ &   $0$         &$\frac{5}{2}$ & $ \frac{1}{2}  $& $     0      $  \\
${\bar N}_H$ &   $0$       &$-\frac{5}{2}$& $-\frac{1}{2}$&$   0      $  \\
$ {\zeta}_H$ &   $0$       & $      0    $& $        1   $& $  1      $  \\
${\bar\zeta}_H$&  $0$      & $      0    $& $       -1   $& $- 1      $  \\
\hline
\end{tabular}
\label{table1}
\eeqn
with $i=1,2,3$. The $U(1)_{\zeta^\prime}$ symmetry is the
combination of $U(1)_{Z^\prime}$ and $U(1)_{\zeta}$ left
unbroken by the vevs of $N_H$ and ${\bar N}_H$.
The fields $\zeta_H$ and ${\bar\zeta}_H$ are needed to break the
residual $U(1)_{\zeta^\prime}$
symmetry. States with the required quantum numbers  
in (\ref{table1}) exist in the string models \cite{lrs}. 
The fields $\phi^i$ are employed in an extended seesaw
mechanism. Using the superpotential terms 
\beq
L_iN_jH^U~~~,~~~N_i{\bar N}_H\phi_j~~~,~~~\phi_i\phi_j\phi_k~.
\eeq
The neutrino seesaw mass matrix takes the form 
\begin{equation}
{\left(\matrix{
                 {\nu_i}&{N_k}&{\phi_m}
                }
   \right)}
  {\left(\matrix{
                 0&(kM_{_U})_{ij}&0\cr
                 (kM_{_U})_{ij}&0&M_{\chi}\cr
                 0&M_{\chi}&O(M_\phi)\cr
                }
   \right)}
  {\left(\matrix{
                 {\nu_j}  \cr
                 {N_l}\cr
                 {\phi_n} \cr
                }
   \right)},
\label{nmm}
\end{equation}
with $M_\chi \sim  \langle {\bar N}_H\rangle$
and  $M_\phi \sim  \langle \phi_0\rangle$.
The mass eigenstates are mainly $\nu_i$, $N_k$
and $\phi_m$ with a small mixing and with the eigenvalues
$$m_{\nu_j} \sim M_\phi \left({{k M^j_u} \over M_{\chi}}\right)^2
\qquad m_{N_j},m_{\phi} \sim M_{\chi}~.$$
A detailed fit to the neutrino data was discussed in ref \cite{nm}.
We emphasize, however, that our aim here is merely to demonstrate that
the extra $U(1)_{\zeta^\prime}$, introduced below, is not in conflict
with the requirement of light neutrino masses.
Alternatively, the VEV of $\langle{\bar N}_H\rangle$ induces heavy Majorana
mass terms for the right--handed neutrinos from nonrenormalizable
terms
\beq
N_iN_j{\bar N}_H{\bar N}_H~.
\eeq
The effective Majorana mass scale of the right--handed neutrinos is then
$M_\chi\sim\langle {\bar N}_H\rangle^2/M$, which for
$\langle{\bar N}_H\rangle\sim10^{16}$GeV gives $M_\chi\sim10^{14}$GeV.
The VEV of $\langle N_H\rangle$ may induce unsuppressed
dimension four baryon and lepton number violating interactions 
\beq
\eta_1QDL+\eta_2UDD
\label{d4op}
\eeq from
the nonrenormalizable terms given in eqs. (\ref{uddn}) and (\ref{qdln}).
Therefore, if the VEV of $N_H$ is of the order of the GUT, or
intermediate, scale, 
as is required in the seesaw mass matrix in eq. (\ref{nmm}),
then unsuppressed proton decay will ensue. However, 
this VEV leaves the unbroken combination of 
$U(1)_{Z^\prime}$ and $U(1)_\zeta$ given by
\beq
U(1)_{\zeta^\prime}={1\over 5} U(1)_{Z^\prime}-U(1)_\zeta~.
\label{u1zetaprime}
\eeq
The induced dimension four lepton number violating operator that arises from
eq. (\ref{qdln}) is invariant under $U(1)_{\zeta^\prime}$, whereas
the induced dimension four baryon number violating operator 
that arises from eq. (\ref{uddn}) is not. Hence, to generate an unsuppressed 
dimension four baryon number violating operator we must break 
also $U(1)_{\zeta^\prime}$. Therefore, if $U(1)_{\zeta^\prime}$ 
remains unbroken down to low energies, it suppresses proton decay from
dimension four operators. Similarly, the dimension five baryon and lepton
number violating operators given in eqs. (\ref{qqql}) and (\ref{uddn})
are not invariant under $U(1)_{\zeta^\prime}$ and hence suppressed
if $U(1)_{\zeta^\prime}$ remains unbroken down to low energies. 

\section{Estimate of the $U(1)_{\zeta^\prime}$ mass scale}

The dimension four and five proton decay mediating operators are 
forbidden by the $U(1)_{Z^\prime}$ and $U(1)_{\zeta}$ gauge symmetries. 
These symmetries are broken by some fields and we can estimate the 
required symmetry breaking scale in order to ensure sufficient
suppression. In turn this will indicate the possible mass scale of the
additional $Z_{\zeta^\prime}$ vector boson, and whether it may  exist
in the range accessible to forthcoming experiments. The dimension 
four operators that give rise to rapid proton decay,
$\eta_1UDD+\eta_2QLD$, are induced from the non--renormalizable 
terms of the form 
\beq
\eta_1(UDDN)\Phi+\eta_2(QLDN)\Phi^\prime
\label{uddnqldnphi}
\eeq
where, $\Phi$ and $\Phi^\prime$ are combinations of fields 
that fix gauge invariance and the string selection rules.
The field $N_H$ can be
the Standard Model singlet in the 16 representation of $SO(10)$, or
it can be a product of two fields, which effectively reproduces 
the $SO(10)$ charges of $N_H$ \cite{ps}. We take the VEV of $N_H$, which 
breaks the $B-L$ symmetry, to be of the order of the GUT scale,
{\it i.e.} $\langle N_H\rangle\sim10^{16}{\rm GeV}$. This is the case
as the VEV of ${\bar N}_H$ induces the seesaw mechanism, which suppresses
the left--handed neutrino masses. The VEVs of $\Phi$ and $\Phi^\prime$
then fixes the magnitude of the effective proton decay mediating 
operators, with 
\beq
\eta_1^\prime\sim{{\langle{N_H}\rangle}\over{M}}
		\left({\langle\phi\rangle\over{M}}\right)^n~~;~~
\eta_2^\prime\sim{{\langle{N_H}\rangle}\over{M}}
\left({\langle{\phi^\prime}\rangle\over{M}}\right)^{n^\prime}.
\label{etaprime}
\eeq

We take $M$ to be the heterotic string unification scale, 
$M\sim10^{18}{\rm GeV}$. 
Similarly, the dimension five proton decay mediating operator
$QQQL$ can effectively be induced from the nonrenormalizable
terms 
\beq
\lambda_1 QQQL(\Phi^{\prime\prime})
\eeq
The VEV of $\phi^{\prime\prime}$ then fixes the
magnitude of the effective dimension five operator to be
\beq
\lambda_1^\prime\sim\lambda_1\left(
{{\langle\phi^{\prime\prime}\rangle}\over{M}}\right)^{n^{\prime\prime}}
\eeq
The experimental limits impose that the product 
$(\eta^\prime_1\eta_2^\prime)\le10^{-24}$ and
$(\lambda_1^\prime/M)\le 10^{-25}$. Hence, for 
$M\sim M_{\rm string}\sim 10^{18}{\rm GeV}$ 
we must have $\lambda_1^\prime\le10^{-7}$, to guarantee that
the proton lifetime is within the experimental bounds.
The induced dimension four lepton number violating operator
is invariant under $U(1)_{\zeta^\prime}$. Hence, we can take
$n^\prime=0$. The dimension five baryon number violating 
operator is not invariant under $U(1)_{\zeta^\prime}$. Hence
we must have at least $n^{\prime\prime}=1$. 
We assume that the dimension four baryon number violating operator
in eq. (\ref{d4op}) is induced at the
quintic order. The corresponding 
nonrenormalizable term in eq. (\ref{uddnqldnphi})
contain one additional field that breaks
the proton protecting $U(1)_{\zeta^\prime}$ at intermediate energy
scale $\Lambda_{\zeta^\prime}$. Hence, we
have $n=1$ in eq. (\ref{etaprime}), and
\beq
({\eta^\prime_1\eta^\prime_2})\sim
\left({{\langle N\rangle}\over{M}}\right)^2
\left({{\Lambda_{\zeta^\prime}}\over{M}}\right)^1
\eeq
Taking $\langle N\rangle\sim 10^{16}{\rm GeV}$ and 
$M\sim 10^{18}{\rm GeV}$, we obtain the estimate 
$\Lambda_{\zeta^\prime}\le 10^{-2}{\rm GeV}$,
which is clearly too low. Taking $\langle N\rangle\sim 10^{13}{\rm GeV}$
yields $\Lambda_{\zeta^\prime}\le 10^{4}{\rm GeV}$.
We also have
that in this case $\lambda_1^\prime/M< 10^{-14}$. Hence, the 
baryon and lepton number violating dimension five operator is adequately 
suppressed. On the other hand, we have $\eta_2^\prime\sim 10^{-5}$. This 
may be too small to produce sizable effects in forthcoming collider
experiments, but may have interesting consequences
for neutralino dark matter searches.

\section{Conclusions} 

The Standard Model gauge and matter spectrum clearly indicates
the realization of grand unification structures in nature. Most appealing in
this respect is the structure of unification in the context of 
embedding the Standard Model chiral spectrum into spinorial
representations of $SO(10)$. In this case each Standard Model 
generation together with the right--handed neutrino fits 
into a single $SO(10)$ spinorial representation. While this
can be a mirage, it is the 
strongest hint from the 
available experimental data, accumulated over the past century.
On the other hand, grand unified theories, and many other extensions
of the renormalizable Standard Model, 
predict processes that lead to proton instability and decay.
Proton longevity is therefore another key ingredient in trying
to understand the fundamental origin of the Standard Model matter
spectrum and interactions. A model that provides
a robust explanation for these two key observations, while
not interfering with other experimental and theoretical constraints,
may indeed stand a good chance to pass further experimental scrutiny. 

String theory provides a viable framework for perturbative 
quantum gravity, while at the same time giving rise to the
gauge and matter structures that describe the interactions of
the Standard Model. In this respect string theory is unique and enables
the development of a phenomenological approach to the unification
of the gauge and gravitational interactions. Heterotic--string theory
has the further distinction that by giving rise to 
spinorial representations in the massless spectrum it also
enables the embedding of the Standard Model chiral spectrum
in $SO(10)$ spinorial representations. 
The free fermionic models provide examples of 
quasi--realistic three generation heterotic--string models,
in which the chiral spectrum arises from $SO(10)$ spinorial
representations. These models therefore admit the $SO(10)$ embedding
of the Standard Model matter states. They satisfy the
two pivotal criteria suggested by the Standard Model data. 
These models are related to $Z_2\times Z_2$ orbifolds at special points
in the moduli space. 
Other classes of quasi--realistic perturbative 
heterotic--string models have also been
studied on unrelated compactifications and using different
techniques \cite{otherclasses}. 

Perhaps the most appealing explanation for the stability
of the proton is the existence of additional gauge symmetries
that forbid the proton decay mediating operators. However,
such gauge symmetries should not interfere, or obstruct, the 
other phenomenological requirements that must be imposed 
on any extension of the Standard Model. Therefore, they should
allow for generation of fermion masses and suppression of neutrino
masses. They should be anomaly free. Gauge symmetries that may 
be observed in forthcoming collider experiments should also
be family universal. 

In this paper we examined the question of such an additional 
$U(1)$ gauge symmetry in the free fermionic models. While
in most cases the additional gauge symmetries that arise
in the string models do not satisfy the needed requirements,
we demonstrated the existence of a $U(1)$ symmetry in the
class of models of ref. \cite{lrs} that indeed does pass all the
criteria. The existence of this $U(1)$ symmetry at low energies
is therefore motivated by the fact that it protects the proton from decaying,
and it may indeed exist in the range accessible to forthcoming
experiments. It is noted that although we investigated the
additional $U(1)$ in the context of the free fermionic string models,
the properties of the $U(1)$ symmetry, and the charges of the Standard
Model state under it, rely solely on the weight charges of the 
string states under the rank 16 gauge symmetry of the ten dimensional
theory. A $U(1)$ symmetry with the properties that we extracted here
may therefore arise in other classes of string compactifications.
We emphasize that the characteristics of the extra $U(1)$ that
we extracted from a particular class of free fermionic models,
do not depend on the specific string compactification.  
It ought to be further noted that compactifications that yielded 
the $U(1)$ and the peculiar Standard Model charges under it, are not 
decedent from the $E_8\times E_8$ heterotic string in 10 dimensions.
This is because a $U(1)$ symmetry which 
descends from the $E_8\times E_8$ (or $SO(16)\times SO(16)$) will 
necessarily have an embedding in $E_6$ and as we demonstrated here
the Standard Model $U(1)$ charges derived in this paper do not possess
an $E_6$ embedding, and do not descend from $E_8$. 
The properties of this $U(1)$ symmetry therefore differ from those
that have been predominantly explored in the literature, which are
inspired from compactifications of the $E_8\times E_8$
heterotic string. The investigation of the phenomenological
characteristics of this additional $U(1)$ is therefore of
further interest and we shall return to it in future publications.

\section{Acknowledgments}

AEF would to thank the 
Oxford theory department for hospitality during the completion of this work. 
CC would like to thank the Liverpool theory division for hospitality. 
This work was supported in part by PPARC (PP/D000416/1), by the Royal Society
and by the Marie Curie Training Research Network ``UniverseNet''
MRTN--CT--2006--035863.

\newpage

\end{document}